\documentclass[reprint,prb,aps,amsmath,amssymb,superscriptaddress,longbibliography,floatfix]{revtex4-2}
\usepackage{graphicx}
\usepackage{bm}
\usepackage{xcolor}
\usepackage[colorlinks=true,linkcolor=blue,citecolor=blue,urlcolor=blue]{hyperref}

\newcommand{\iu}{\mathrm{i}}
\newcommand{\dd}{\mathrm{d}}
\newcommand{\Tr}{\mathrm{Tr}}

\newcommand{\GB}{\mathbf{G}^{B}}
\newcommand{\Kv}{\boldsymbol{\mathcal{K}}}
\newcommand{\Dani}{\boldsymbol{\Delta}}

\begin{document}
\raggedbottom

\title{Zeeman-field-induced three-spin interactions and relativistic coupling to the scalar spin chirality}

\author{Samir Lounis}
\email{samir.lounis@physik.uni-halle.de}
\affiliation{Institute of Physics and Halle-Berlin-Regensburg Cluster of Excellence CCE, Martin Luther University Halle-Wittenberg, 06120 Halle (Saale), Germany}

\date{\today}

\begin{abstract}
Magnetic fields are commonly incorporated in atomistic spin models via the Zeeman energy of the individual magnetic moments. Here, we demonstrate, starting from a microscopic electronic Hamiltonian, that the Zeeman coupling of the conduction electrons generates in addition three-spin interactions linear in the field, which are the lowest-order corrections to the Zeeman energy allowed by time-reversal symmetry. Utilizing a Green-function expansion of the grand potential, we identify two classes of such interactions. The isotropic ones, of the form $(\mathbf{S}_i\cdot\mathbf{S}_j)(\mathbf{S}_k\cdot\mathbf{B})$, exist without spin-orbit coupling, and their total strength obeys a sum rule relating it to the three-site electronic loop at the Fermi energy. The relativistic ones require spin-orbit coupling and include a coupling $\boldsymbol{\mathcal{K}}_{ijk}\cdot\mathbf{B}\,\chi_{ijk}$ between the field and the scalar spin chirality $\chi_{ijk}=\mathbf{S}_i\cdot(\mathbf{S}_j\times\mathbf{S}_k)$. We show that such a coupling is forbidden in the spin channel without spin-orbit coupling, in contrast to the orbital channel, which we treat within the same framework and which recovers the orbital-flux mechanism of Mott insulators as the Zeeman energy of the chirality-induced loop current. Exact diagonalization of the Hubbard trimer confirms both channels. For magnetic atoms embedded in one- and two-dimensional Rashba electron gases, we derive closed expressions to leading order in the spin-orbit coupling: $\boldsymbol{\mathcal{K}}_{ijk}$ lies in-plane at linear order in the Rashba coupling and cancels for $C_{3v}$-symmetric trimers, for which an out-of-plane component emerges at second order. In magnetic trimers, the three-spin terms shift the critical fields of magnetization steps and modify the field-induced non-coplanarity. In the continuum limit, the chirality coupling becomes an energy proportional to the topological charge, which splits skyrmions from antiskyrmions, while the relativistic terms produce a field-dependent Dzyaloshinskii-Moriya interaction.
\end{abstract}

\maketitle

\section{Introduction}
\label{sec:intro}

Non-collinear spin-textures, such as spin spirals, domain walls, magnetic skyrmions~\cite{Fert2017,Nagaosa2013}, hedgehog lattices~\cite{kanazawa_critical_2016} or hopfions~\cite{Wang2019}, are at the heart of intensive investigations, and atomistic spin models constitute the natural framework to describe them. Their occurrence is driven by competing magnetic exchange interactions (MEI): the bilinear isotropic exchange~\cite{Heisenberg1928,Anderson1959}, the Dzyaloshinskii-Moriya interaction (DMI)~\cite{Dzyaloshinskii1958,Moriya1960} and the symmetric anisotropic exchange~\cite{vanVleck1937,Moriya1953}. It is by now established that higher-order multi-spin interactions can be equally important in many materials~\cite{Takahashi1977,MacDonald1988,Roger1983,Heinze2011,Paul2020,Hoffmann2020}. These comprise the isotropic biquadratic and four-spin interactions~\cite{Kittel1960,Harris1963,Huang1964,Uryu1965,Iwashita1974} as well as their chiral counterparts~\cite{Brinker2019,Laszloffy2019,Grytsiuk2020,Brinker2020,dosSantosDias2021}, whose microscopic origin was unveiled within multiple-scattering theory~\cite{Brinker2019,Grytsiuk2020,Brinker2020,Lounis2020}. In itinerant magnets, such interactions can stabilize multiple-$Q$ states and topological spin-textures~\cite{Shindou2001,Heinze2011,Martin2008,Akagi2010,Batista2016,Ozawa2017,Hayami2017,Hayami2021,Okumura2020}.

Interestingly, all these interactions involve an even number of spins. This is a direct consequence of time-reversal symmetry $\mathcal{T}$: every spin changes sign under $\mathcal{T}$, and a product of $n$ spins is $\mathcal{T}$-invariant only if $n$ is even. The situation changes once an external magnetic field $\mathbf{B}$ is applied, since $\mathbf{B}$ is itself $\mathcal{T}$-odd, and terms with an odd number of spins multiplied by an odd power of $\mathbf{B}$ become allowed. To linear order in the field, the one-spin term is nothing else than the familiar Zeeman energy $-\mathbf{M}_i\cdot\mathbf{B}$, while the next allowed terms involve three spins. Symmetry allows such terms but does not guarantee their existence. In SU(2)-symmetric Mott insulators, for instance, they are absent: the Zeeman coupling acts on the total spin, which commutes with the Hubbard Hamiltonian, so that the field enters the low-energy spin model exactly as $-\mathbf{B}\cdot\sum_i\mathbf{S}_i$, and the $t/U$ expansion~\cite{Takahashi1977,MacDonald1988} does not generate any field-induced multi-spin term. Field-induced multi-spin interactions require therefore either spin-orbit coupling (SOC), which breaks spin-rotation invariance, or a situation where the field acts on electrons whose spin is not locked to the local moments. An example of the first case is the three-spin term that a magnetic field generates at third order in Kitaev's honeycomb model, where the SOC-derived bond-dependent exchange lifts the spin-rotation symmetry~\cite{Kitaev2006}. The second case is generic in metallic magnets, where the field polarizes the conduction electrons mediating the interactions between the moments.

While the even multi-spin interactions of itinerant magnets have been addressed in great detail, the field-linear three-spin terms received little attention. A field-linear coupling to the scalar chirality is anticipated from the orbital moment induced by non-coplanar spins~\cite{Shindou2001,Dias2016,Hanke2016}, whose Zeeman energy is linear in the field. In Mott insulators, it corresponds to the orbital-flux mechanism discussed below~\cite{Sen1995,Chitra2001}. In these works, the coupling originates from the orbital motion of the electrons and involves no SOC. Whether, and under which conditions, the chirality can also couple to the field through the electron spin was however not addressed, and as we demonstrate in the following, this requires SOC. A field-induced chiral three-spin interaction was also derived for Weyl electrons~\cite{Nikolic2021}. Nevertheless, a general microscopic theory of all field-linear three-spin terms, in particular of those generated by the Zeeman coupling of the electron spin, is lacking, and in practice spin models of metallic magnets are combined with the bare Zeeman energy only.

Among the three-spin quantities, the scalar spin chirality $\chi_{ijk}=\mathbf{S}_i\cdot(\mathbf{S}_j\times\mathbf{S}_k)$ plays a special role. It measures the solid angle spanned by three non-coplanar spins and is at the origin of the topological Hall effect~\cite{Taguchi2001,Bruno2004,Neubauer2009}, of the chiral orbital magnetism~\cite{Hoffmann2015,Dias2016,Hanke2016,Dias2017} and of chiral spin liquids~\cite{Kalmeyer1987,Wen1989}. Being $\mathcal{T}$-odd, $\chi_{ijk}$ cannot appear alone in a zero-field spin Hamiltonian. Without field, it shows up only in combination with an additional spin, as in the four-spin interaction $(\boldsymbol{\mathcal{C}}_{ijkl}\cdot\mathbf{S}_l)\,\chi_{ijk}$~\cite{Grytsiuk2020,Brinker2020}, or via the six-spin product $\chi_{ijk}\chi_{lmn}$~\cite{Grytsiuk2020}. In Mott insulators, a three-spin term $\propto\chi_{ijk}$ emerges at third order in the hopping from the \emph{orbital} effect of a magnetic field, i.e., from the flux threading the triangle $ijk$~\cite{Sen1995,Chitra2001,Motrunich2006}, and such a term can stabilize a chiral spin liquid~\cite{Bauer2014}. This mechanism does not involve SOC. For the Zeeman coupling of the electron spin, a symmetry argument (Sec.~\ref{sec:symmetry}) shows that such a term requires SOC, and it was obtained for Weyl electrons, whose spin is locked to their momentum~\cite{Nikolic2021}. A microscopic theory valid for arbitrary hosts, which determines the origin, magnitude, direction and geometry dependence of this coupling, is however missing.

In this work, we derive analytically the field-induced three-spin interactions starting from a microscopic electronic Hamiltonian. Utilizing a Green-function expansion of the grand potential, in the spirit of the infinitesimal-rotation method~\cite{Liechtenstein1987,Brinker2019,Grytsiuk2020,Lounis2020}, we identify two classes of interactions. The first one is isotropic, of the form $\mathcal{Z}_{i(jk)}(\mathbf{S}_j\cdot\mathbf{S}_k)(\mathbf{S}_i\cdot\mathbf{B})$, and exists without SOC. It originates from the field-induced spin polarization of the conduction electrons, has no counterpart in SU(2)-symmetric Mott insulators, and its total strength obeys a simple Fermi-surface sum rule. The second class is relativistic and contains in particular the term $\boldsymbol{\mathcal{K}}_{ijk}\cdot\mathbf{B}\,\chi_{ijk}$, where the vector $\boldsymbol{\mathcal{K}}_{ijk}$ plays a role similar to that of the DMI vector. The derivation unveils how this chirality coupling emerges from scattering processes in which one electron propagator carries the spin dependence induced by SOC, and provides an explicit expression for $\boldsymbol{\mathcal{K}}_{ijk}$ in terms of the electronic Green functions. In line with symmetry, $\boldsymbol{\mathcal{K}}_{ijk}$ vanishes identically without SOC. We then evaluate all coefficients for magnetic atoms embedded in one- (1D) and two-dimensional (2D) Rashba electron gases~\cite{Rashba1960,Bychkov1984}. The resulting closed expressions reveal features that cannot be anticipated from symmetry alone. For instance, the in-plane chirality coupling cancels in $C_{3v}$-symmetric trimers at linear order in the Rashba coupling, while it survives in linear trimers. Finally, we address the consequences for magnetic trimers, which can be built and probed atom by atom with scanning tunneling microscopy~\cite{Wahl2007,Meier:2008,Zhou2010,Khajetoorians2012,Khajetoorians2016}.

The manuscript is organized as follows. After a symmetry analysis in Sec.~\ref{sec:symmetry}, we present the microscopic derivation in Sec.~\ref{sec:theory}. Section~\ref{sec:orbital} deals with the orbital coupling of the field and compares both channels in the Mott limit, while Sec.~\ref{sec:rashba} is devoted to the Rashba model. The impact on magnetic trimers and the implications for skyrmions are discussed in Sec.~\ref{sec:trimers}, before we conclude in Sec.~\ref{sec:conclusion}. Technical details are collected in the Appendices.

\section{Symmetry considerations}
\label{sec:symmetry}

We are looking for energy contributions that are linear in $\mathbf{B}$ and cubic in the unit vectors $\mathbf{S}_i$, $\mathbf{S}_j$, $\mathbf{S}_k$ of three distinct magnetic moments, with $\mathbf{S}=\mathbf{M}/|\mathbf{M}|$. Such terms are automatically $\mathcal{T}$-even. An additional constraint comes from spin-rotation symmetry. Without SOC, spin space is decoupled from real space, and the energy must be invariant under a global rotation of all spins together with the field. The only scalars linear in each of the four vectors are then products of two scalar products, since contractions with the Levi-Civita symbol reduce to scalar products via $\epsilon_{abc}\epsilon_{ade}=\delta_{bd}\delta_{ce}-\delta_{be}\delta_{cd}$. Hence, without SOC, the only allowed field-linear terms of third order in the spins are
\begin{equation}
(\mathbf{S}_j\cdot\mathbf{S}_k)(\mathbf{S}_i\cdot\mathbf{B}), \quad
(\mathbf{S}_k\cdot\mathbf{S}_i)(\mathbf{S}_j\cdot\mathbf{B}), \quad
(\mathbf{S}_i\cdot\mathbf{S}_j)(\mathbf{S}_k\cdot\mathbf{B}).
\label{eq:iso_invariants}
\end{equation}
The product $\chi_{ijk}\,\mathbf{B}$, however, is a vector in spin space, which has to be contracted with a second vector to give an energy. In the spin channel, such a vector can only be provided by the lattice via SOC, which locks spin space to real space. This leads to terms of the form $\boldsymbol{\mathcal{K}}_{ijk}\cdot\mathbf{B}\,\chi_{ijk}$ and to further anisotropic terms such as $\boldsymbol{\mathcal{K}}'\cdot(\mathbf{S}_i\times\mathbf{B})(\mathbf{S}_j\cdot\mathbf{S}_k)$, all of which require SOC. The orbital mechanism of Refs.~\cite{Sen1995,Motrunich2006,Bauer2014} escapes this argument, since the orbital field enters through the flux $\Phi_{ijk}$ threading the triangle, which is invariant under spin rotations. These conclusions are summarized in Table~\ref{tab:symmetry}. Note that symmetry only allows these terms. Whether they appear, and with which strength, is a question for the microscopic theory, and, as aforementioned, the isotropic terms vanish identically in SU(2)-symmetric Mott insulators.

Similarly to Moriya's rules for the DMI~\cite{Moriya1960}, the direction of $\boldsymbol{\mathcal{K}}_{ijk}$ is dictated by the point-group symmetry of the cluster. Two cases are of interest in the following. For an equilateral trimer with $C_{3v}$ symmetry, invariance under the threefold axis enforces $\boldsymbol{\mathcal{K}}_{ijk}\parallel\hat{z}$, where $\hat{z}$ is the surface normal, and one can verify that $B_z\chi_{ijk}$ is also invariant under the vertical mirror planes. For a linear trimer along $\hat{x}$ on a surface, the two mirror planes, i.e., the $xz$ plane containing the chain and the $yz$ plane through the central atom, allow only $\boldsymbol{\mathcal{K}}_{ijk}\parallel\hat{y}$: the vector lies in-plane and is perpendicular to the chain.

\begin{table}[b]
\caption{Field-linear energy terms involving at most three spins. $\mathcal{T}$: even under time reversal; rot.: invariant under a global rotation of spins and field in the absence of SOC; spin/orbital: coupling of the field to the electron spin or orbital motion.}
\label{tab:symmetry}
\begin{ruledtabular}
\begin{tabular}{lccl}
Term & $\mathcal{T}$ & rot. & Origin \\
\hline
$-\mathbf{M}_i\cdot\mathbf{B}$ & yes & yes & Zeeman \\
$(\mathbf{S}_j\cdot\mathbf{S}_k)(\mathbf{S}_i\cdot\mathbf{B})$ & yes & yes & spin, no SOC \\
$\Phi_{ijk}\,\chi_{ijk}$ & yes & yes & orbital~\cite{Sen1995,Motrunich2006} \\
$\boldsymbol{\mathcal{K}}_{ijk}\cdot\mathbf{B}\,\chi_{ijk}$ & yes & no & spin, needs SOC \\
$\propto\mathbf{B}$, two spins & no & -- & forbidden \\
\end{tabular}
\end{ruledtabular}
\end{table}

\section{Microscopic theory}
\label{sec:theory}

\subsection{Model and expansion of the grand potential}

We consider the electronic Hamiltonian
\begin{equation}
\mathcal{H} = \mathcal{H}^{0} + \mathcal{H}^{\mathrm{soc}} + \mathcal{H}^{\mathrm{mag}} + \mathcal{H}^{\mathrm{Z}},
\label{eq:hamiltonian}
\end{equation}
where $\mathcal{H}^{0}$ contains all spin-independent terms, $\mathcal{H}^{\mathrm{soc}}$ is the SOC, and $\mathcal{H}^{\mathrm{mag}}=\sum_i U_i\,\mathbf{S}_i\cdot\boldsymbol{\sigma}$ describes the local exchange coupling of strength $U_i$ between the moment direction $\mathbf{S}_i$ and the electron spin, $\boldsymbol{\sigma}$ being the vector of Pauli matrices. The Zeeman coupling of the electrons, $\mathcal{H}^{\mathrm{Z}}=-\mathbf{B}\cdot\boldsymbol{\sigma}$, acts on every site of the system. Throughout the paper, $\mathbf{B}$ denotes the Zeeman energy, i.e., the factor $g\mu_B/2$ is absorbed in $\mathbf{B}$, and the moments $\mathbf{M}_i$ are measured in units of $2/g$ times the spin moment, such that the Zeeman energy of a single moment reads $-\mathbf{M}_i\cdot\mathbf{B}$.

The magnetic interactions are identified from the change of the grand potential upon a perturbation $\delta\mathcal{H}$ of a reference Hamiltonian $\mathcal{H}^{\mathrm{ini}}$. Utilizing Lloyd's formula, one obtains the formal power series~\cite{Lloyd1972,Liechtenstein1987,Brinker2019,Lounis2020}
\begin{equation}
\Omega = \Omega^{\mathrm{ini}} - \frac{1}{\pi}\,\mathrm{Im}\!\int\!\dd E\, f(E)\sum_{p\geq1}\frac{1}{p}\,\Tr\left[G(E)\,\delta\mathcal{H}\right]^{p},
\label{eq:lloyd}
\end{equation}
where $G(E)=(E+\iu0^{+}-\mathcal{H}^{\mathrm{ini}})^{-1}$ is the retarded Green function of the reference system, $f(E)$ the Fermi-Dirac distribution, and the trace runs over sites, orbitals and spin. Here, $\mathcal{H}^{\mathrm{ini}}=\mathcal{H}^{0}+\mathcal{H}^{\mathrm{soc}}$, while $\delta\mathcal{H}=\mathcal{H}^{\mathrm{mag}}+\mathcal{H}^{\mathrm{Z}}$ is expanded to first order in $\mathbf{B}$ and to third order in the spins. Following the infinitesimal-rotation approach~\cite{Liechtenstein1987,Brinker2019,Grytsiuk2020}, the obtained coefficients are identified with those of the corresponding spin Hamiltonian. We note that the coefficients given below are of leading order in the exchange coupling $U_i$. Replacing $U_i$ by the scattering matrices $t_i$ of the magnetic atoms resums the multiple scattering at each site and extends them beyond weak coupling~\cite{Brinker2019,Lounis2020}, while the Rashba evaluation of Sec.~\ref{sec:rashba} relies on the lowest-order (Born) approximation. An exact test on a tight-binding metal (Appendix~\ref{app:exact}) shows that the leading-order coefficients are accurate within a few percent for $U_i\lesssim t/2$ and within 8\% for $U_i=t$, i.e., up to about one eighth of the bandwidth.

\begin{figure}[t]
\includegraphics[width=\linewidth]{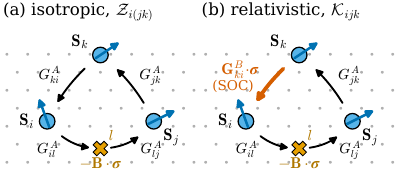}
\caption{Scattering processes generating the field-induced three-spin interactions. Conduction electrons of the host (gray lattice) propagate between the magnetic atoms $i$, $j$ and $k$, whose moments $\mathbf{S}$ (arrows) couple to the electron spin via $U\,\mathbf{S}\cdot\boldsymbol{\sigma}$, and interact with the field at an arbitrary site $l$ (cross), which is summed over. (a) Isotropic process involving only the spin-independent propagator $G^{A}$, which yields the coefficients $\mathcal{Z}_{i(jk)}$. (b) Relativistic process, in which one propagator is replaced by its SOC-induced spin-dependent part $\mathbf{G}^{B}\cdot\boldsymbol{\sigma}$. Shown here on bond $ki$, it can equally sit on $jk$, $il$ or $lj$. These processes yield the chirality coupling $\boldsymbol{\mathcal{K}}_{ijk}$ and the other relativistic terms.}
\label{fig:processes}
\end{figure}

In spin space, the Green function between sites $i$ and $j$ decomposes as
\begin{equation}
G_{ij}(E) = G^{A}_{ij}(E)\,\sigma_0 + \mathbf{G}^{B}_{ij}(E)\cdot\boldsymbol{\sigma}.
\label{eq:green_decomposition}
\end{equation}
For a non-magnetic reference system, $\mathbf{G}^{B}_{ij}$ originates entirely from SOC. Time-reversal symmetry of the reference system implies $G_{ij}=\sigma_y G_{ji}^{\mathsf{T}}\sigma_y$, and hence $G^{A}_{ij}=G^{A}_{ji}$ and $\mathbf{G}^{B}_{ij}=-\mathbf{G}^{B}_{ji}$ (up to a transposition of the orbital indices). If, in addition, the bond $ij$ has an inversion center, $\mathbf{G}^{B}_{ij}$ vanishes, in close analogy to Moriya's rules. Since the reference is time-reversal invariant, the orbital trace of $\mathbf{G}^{B}_{ii}$ vanishes as well, and the first-order term of Eq.~\eqref{eq:lloyd} with $\delta\mathcal{H}=\mathcal{H}^{\mathrm{Z}}$ does not contribute. The conventional Zeeman energy emerges at second order, from one Zeeman vertex and one exchange vertex:
\begin{align}
\Omega^{(2)}_{\mathrm{Z}} &= \frac{2}{\pi}\,\mathrm{Im}\!\int\!\dd E\,f(E)\sum_{i,l}\Tr_{o}\big[G^{A}_{il}U_iG^{A}_{li}\big]\,\mathbf{S}_i\cdot\mathbf{B} \nonumber\\
&= -\sum_i M_i\,\mathbf{S}_i\cdot\mathbf{B},
\label{eq:zeeman}
\end{align}
where $\Tr_o$ is the trace over orbitals and $M_i$ is the spin polarization of the electrons induced by the moment at site $i$. In a rigid-spin description, $M_i$ includes the intrinsic moment of the magnetic atom. The three-spin terms arise at fourth order, $p=4$, from products containing $\mathcal{H}^{\mathrm{Z}}$ once and $\mathcal{H}^{\mathrm{mag}}$ on three distinct sites $i$, $j$ and $k$. The Zeeman vertex sits on an arbitrary site $l$, which is summed over the whole system, as illustrated in Fig.~\ref{fig:processes}. The four cyclic positions of $\mathcal{H}^{\mathrm{Z}}$ in the trace cancel the prefactor $1/p$, and the two orientations of the loop $i\to j\to k\to i$ contribute equally. We verified the latter property explicitly using $G^{A}_{ij}=G^{A}_{ji}$, $\mathbf{G}^{B}_{ij}=-\mathbf{G}^{B}_{ji}$ and the Pauli-trace identities of Appendix~\ref{app:traces}.

\subsection{Isotropic three-spin interactions}

Retaining only $G^{A}$ in all four propagators, the spin trace $\Tr_s[(\boldsymbol{\sigma}\cdot\mathbf{S}_j)(\boldsymbol{\sigma}\cdot\mathbf{S}_k)(\boldsymbol{\sigma}\cdot\mathbf{S}_i)(\boldsymbol{\sigma}\cdot\mathbf{B})]$ generates exactly the three invariants of Eq.~\eqref{eq:iso_invariants}, as expected from symmetry. The contribution of a trimer reads
\begin{align}
\Omega^{\mathrm{iso}}_{ijk} &= \mathcal{Z}_{i(jk)}\,(\mathbf{S}_j\cdot\mathbf{S}_k)(\mathbf{S}_i\cdot\mathbf{B}) + \mathcal{Z}_{j(ki)}\,(\mathbf{S}_k\cdot\mathbf{S}_i)(\mathbf{S}_j\cdot\mathbf{B}) \nonumber\\
&\quad + \mathcal{Z}_{k(ij)}\,(\mathbf{S}_i\cdot\mathbf{S}_j)(\mathbf{S}_k\cdot\mathbf{B}),
\label{eq:omega_iso}
\end{align}
where the subscript $i(jk)$ indicates that the field couples to $\mathbf{S}_i$, while $\mathbf{S}_j$ and $\mathbf{S}_k$ enter via their scalar product. The coefficients are given by
\begin{equation}
\mathcal{Z}_{i(jk)} = 4\left(\Delta_{\overline{ij}k} + \Delta_{\overline{ki}j} - \Delta_{\overline{jk}i}\right),
\label{eq:Z_iso}
\end{equation}
with
\begin{align}
\Delta_{\overline{ij}k} &= \frac{1}{\pi}\,\mathrm{Im}\!\int\!\dd E\, f(E)\sum_l \Tr_o\left[G^{A}_{il}G^{A}_{lj}U_jG^{A}_{jk}U_kG^{A}_{ki}U_i\right] \nonumber\\
&= -\frac{1}{\pi}\,\mathrm{Im}\!\int\!\dd E\, f(E)\,\Tr_o\left[\frac{\dd G^{A}_{ij}}{\dd E}U_jG^{A}_{jk}U_kG^{A}_{ki}U_i\right].
\label{eq:Delta_iso}
\end{align}
The bar marks the bond $ij$ on which the field vertex is inserted. In the second equality, we used $\sum_l G_{il}G_{lj}=-\dd G_{ij}/\dd E$, which holds for $G^{A}$ up to a correction of second order in $\mathbf{G}^{B}$. The physical picture is simple: the field shifts the Fermi levels of the two spin channels in opposite directions, and $\Delta_{\overline{ij}k}$ describes the resulting change of the phase accumulated along bond $ij$ during a three-site loop. Interestingly, the bond whose two spins are contracted in a given term, $jk$ for $\mathcal{Z}_{i(jk)}$, enters Eq.~\eqref{eq:Z_iso} with a negative sign.

Summing Eq.~\eqref{eq:Delta_iso} over the three bonds converts the integrand into a total energy derivative, which leads to the sum rule
\begin{align}
\mathcal{X}_{ijk} &\equiv \mathcal{Z}_{i(jk)} + \mathcal{Z}_{j(ki)} + \mathcal{Z}_{k(ij)} \nonumber\\
&= 4\left(\Delta_{\overline{ij}k}+\Delta_{\overline{jk}i}+\Delta_{\overline{ki}j}\right) \nonumber\\
&= -\frac{4}{\pi}\,\mathrm{Im}\!\int\!\dd E\left(-\frac{\partial f}{\partial E}\right)\Tr_o\left[U_iG^{A}_{ij}U_jG^{A}_{jk}U_kG^{A}_{ki}\right],
\label{eq:sum_rule}
\end{align}
which reduces at zero temperature to the three-site loop evaluated at the Fermi energy $E_F$. In other words, the total isotropic three-spin Zeeman coupling of a trimer is a Fermi-surface property, similarly to the Pauli susceptibility, while the individual coefficients $\mathcal{Z}_{i(jk)}$ depend on how the field-induced phase is distributed over the three bonds. Equations~\eqref{eq:Z_iso} and \eqref{eq:sum_rule} agree with exact Lloyd-formula calculations for a metallic host to better than $10^{-3}$ (Appendix~\ref{app:exact}). For collinear configurations, $\mathbf{S}_n=s_n\hat{\mathbf{e}}$ with $s_n=\pm1$ and $\mathbf{B}=B\hat{\mathbf{e}}$, Eq.~\eqref{eq:omega_iso} simplifies to
\begin{equation}
\Omega^{\mathrm{iso}}_{ijk} = \mathcal{X}_{ijk}\,B\,s_i s_j s_k .
\label{eq:ising}
\end{equation}
The three-spin term thus distinguishes configurations with the same total moment but with a different parity of the number of reversed spins, which is impossible with the conventional Zeeman energy.

\subsection{Relativistic three-spin interactions and the scalar spin chirality}

Replacing one of the four propagators by $\mathbf{G}^{B}\cdot\boldsymbol{\sigma}$ leads to a trace of five Pauli matrices, which equals $2\iu$ times a real multilinear form, so that the relativistic coefficients involve the real part of the energy integral. For the loop with the field vertex on bond $ij$ (Fig.~\ref{fig:processes}), $\mathbf{G}^{B}$ can be placed at four positions: on the two bonds $ki$ and $jk$, which do not carry the field, and on either of the two segments $il$ and $lj$ adjacent to the field vertex. We define
\begin{widetext}
\begin{align}
\Dani^{ki}_{\overline{ij}k} &= \frac{1}{\pi}\,\mathrm{Re}\!\int\!\dd E\,f(E)\sum_l\Tr_o\left[G^{A}_{il}G^{A}_{lj}U_j\,G^{A}_{jk}U_k\,\GB_{ki}U_i\right], \nonumber\\
\Dani^{jk}_{\overline{ij}k} &= \frac{1}{\pi}\,\mathrm{Re}\!\int\!\dd E\,f(E)\sum_l\Tr_o\left[G^{A}_{il}G^{A}_{lj}U_j\,\GB_{jk}U_k\,G^{A}_{ki}U_i\right], \nonumber\\
\Dani^{lj}_{\overline{ij}k} &= \frac{1}{\pi}\,\mathrm{Re}\!\int\!\dd E\,f(E)\sum_l\Tr_o\left[G^{A}_{il}\GB_{lj}U_j\,G^{A}_{jk}U_k\,G^{A}_{ki}U_i\right], \nonumber\\
\Dani^{il}_{\overline{ij}k} &= \frac{1}{\pi}\,\mathrm{Re}\!\int\!\dd E\,f(E)\sum_l\Tr_o\left[\GB_{il}G^{A}_{lj}U_j\,G^{A}_{jk}U_k\,G^{A}_{ki}U_i\right].
\label{eq:Delta_ani}
\end{align}
The corresponding contribution to the grand potential is
\begin{align}
\Omega^{\mathrm{ani}}_{\overline{ij}k} = 4\,\Big\{\;
&\Dani^{ki}_{\overline{ij}k}\cdot\left[-\chi_{ijk}\mathbf{B} + (\mathbf{S}_i\times\mathbf{B})(\mathbf{S}_j\cdot\mathbf{S}_k) + (\mathbf{S}_j\times\mathbf{S}_k)(\mathbf{S}_i\cdot\mathbf{B}) + \mathbf{S}_i\,(\mathbf{S}_j\times\mathbf{S}_k)\cdot\mathbf{B}\right] \nonumber\\
+\,&\Dani^{jk}_{\overline{ij}k}\cdot\left[-\chi_{ijk}\mathbf{B} - (\mathbf{S}_j\times\mathbf{B})(\mathbf{S}_k\cdot\mathbf{S}_i) + (\mathbf{S}_k\times\mathbf{S}_i)(\mathbf{S}_j\cdot\mathbf{B}) + \mathbf{S}_j\,(\mathbf{S}_k\times\mathbf{S}_i)\cdot\mathbf{B}\right] \nonumber\\
+\,&\Dani^{lj}_{\overline{ij}k}\cdot\left[+\chi_{ijk}\mathbf{B} + (\mathbf{S}_j\times\mathbf{B})(\mathbf{S}_k\cdot\mathbf{S}_i) - (\mathbf{S}_k\times\mathbf{S}_i)(\mathbf{S}_j\cdot\mathbf{B}) + \mathbf{S}_j\,(\mathbf{S}_k\times\mathbf{S}_i)\cdot\mathbf{B}\right] \nonumber\\
+\,&\Dani^{il}_{\overline{ij}k}\cdot\left[+\chi_{ijk}\mathbf{B} - (\mathbf{S}_i\times\mathbf{B})(\mathbf{S}_j\cdot\mathbf{S}_k) - (\mathbf{S}_j\times\mathbf{S}_k)(\mathbf{S}_i\cdot\mathbf{B}) + \mathbf{S}_i\,(\mathbf{S}_j\times\mathbf{S}_k)\cdot\mathbf{B}\right]\Big\},
\label{eq:omega_ani}
\end{align}
\end{widetext}
and the relativistic energy of the trimer, to linear order in $\mathbf{G}^{B}$, is $\Omega^{\mathrm{ani}}_{ijk}=\Omega^{\mathrm{ani}}_{\overline{ij}k}+\Omega^{\mathrm{ani}}_{\overline{jk}i}+\Omega^{\mathrm{ani}}_{\overline{ki}j}$. The brackets in Eq.~\eqref{eq:omega_ani} follow from the identities listed in Appendix~\ref{app:traces}, each of them being checked numerically.

It is worth noting that the terms explicitly proportional to $\chi_{ijk}\mathbf{B}$ in Eq.~\eqref{eq:omega_ani} do not exhaust the coupling to the chirality. Indeed, the brackets also contain the products $(\mathbf{S}_j\times\mathbf{S}_k)(\mathbf{S}_i\cdot\mathbf{B})$ and $\mathbf{S}_i\,(\mathbf{S}_j\times\mathbf{S}_k)\cdot\mathbf{B}$, each of which hides a component $\frac{1}{3}\chi_{ijk}\mathbf{B}$. We therefore define the chirality coupling as the part of $\Omega^{\mathrm{ani}}_{ijk}$ that is invariant under a global rotation of the spins at fixed $\mathbf{B}$ and fixed SOC vectors. This part is unique and is obtained by averaging over such rotations (Appendix~\ref{app:traces}), which gives
\begin{align}
\Omega^{\chi}_{ijk} &= \Kv_{ijk}\cdot\mathbf{B}\;\chi_{ijk}, \nonumber\\
\Kv_{ijk} &= 4\sum_{\mathrm{cyc}} \Big[\Dani^{lj}_{\overline{ij}k}+\Dani^{il}_{\overline{ij}k}-\tfrac{1}{3}\big(\Dani^{ki}_{\overline{ij}k}+\Dani^{jk}_{\overline{ij}k}\big)\Big],
\label{eq:K}
\end{align}
where the sum runs over the three cyclic permutations of $(ijk)$, which leave $\chi_{ijk}$ invariant. The remaining, non-invariant parts of Eq.~\eqref{eq:omega_ani} depend on the orientation of the spins with respect to the SOC vectors. An exact calculation with Rashba SOC confirms Eq.~\eqref{eq:K}, including the weights $-\frac{1}{3}$, within 3\% at weak SOC (Appendix~\ref{app:exact}). Equation~\eqref{eq:K} is the central result of this work: the scalar spin chirality couples to the Zeeman field at the three-spin level via the SOC-induced vector $\Kv_{ijk}$, which vanishes identically when $\mathbf{G}^{B}=0$. Moreover, the segments $il$ and $lj$ enter $\Kv_{ijk}$ only via their sum, which can again be written as an energy derivative, $\sum_l(G^{A}_{il}\GB_{lj}+\GB_{il}G^{A}_{lj})=-\dd\GB_{ij}/\dd E$, up to corrections of higher order in SOC. Hence
\begin{align}
\Dani^{lj}_{\overline{ij}k}&+\Dani^{il}_{\overline{ij}k} \nonumber\\
&= -\frac{1}{\pi}\,\mathrm{Re}\!\int\!\dd E\,f(E)\,\Tr_o\left[\frac{\dd\GB_{ij}}{\dd E}U_jG^{A}_{jk}U_kG^{A}_{ki}U_i\right].
\label{eq:derivative_B}
\end{align}
Consequently, all contributions to $\Kv_{ijk}$ share a common structure: the field-induced phase acquired on one bond $b$ multiplies a spin-orbit vector originating from a bond $b'$, with weight $+1$ for $b=b'$ and $-\frac{1}{3}$ for $b\neq b'$. The remaining terms of Eq.~\eqref{eq:omega_ani} couple the field to mixed vector and scalar products of the spins. Similarly to the DMI, they favor a chiral canting of the individual moments away from the field, with a sense of rotation set by the vectors $\Dani$.

\subsection{Physical interpretation: configuration-dependent magnetic moment}
\label{sec:interpretation}

The field-induced interactions can be given a simple physical meaning. To linear order in $\mathbf{B}$, the Hellmann-Feynman theorem leads to
\begin{equation}
\Omega(\{\mathbf{S}\},\mathbf{B}) = \Omega_0(\{\mathbf{S}\}) - \mathbf{B}\cdot\boldsymbol{\mu}_{\mathrm{tot}}(\{\mathbf{S}\}) + \mathcal{O}(B^{2}),
\label{eq:hellmann_feynman}
\end{equation}
where $\boldsymbol{\mu}_{\mathrm{tot}}(\{\mathbf{S}\})$ is the total magnetic moment of the magnetic atoms and of the host in the zero-field state with spin configuration $\{\mathbf{S}\}$. The field couples thus to a single quantity, which is, however, not the sum of rigid local moments: it depends on the relative orientation of the moments through the polarization of the host electrons. Expanding $\boldsymbol{\mu}_{\mathrm{tot}}$ in the spins, the linear terms correspond to the rigid moments entering the conventional Zeeman energy, Eq.~\eqref{eq:zeeman}, while the cubic terms of the spin moment are the interactions derived here. $\mathcal{Z}_{i(jk)}$ describes the part of the host spin polarization that depends on the relative orientation of three moments, and $-\Kv_{ijk}\chi_{ijk}$ is a spin moment induced by the scalar chirality. The origin of the latter can be understood as follows. The non-coplanarity of the moments generates ground-state currents, i.e., the chiral orbital moment, even without SOC~\cite{Dias2016,Hanke2016}. Since the Zeeman field acts on the electron spin only, the chirality can couple to the field in this channel only once SOC converts these currents into a spin polarization. For Rashba SOC, a current $\mathbf{j}$ produces a spin polarization $\propto\hat{z}\times\mathbf{j}$, which averages to zero for a current circulating uniformly around a closed loop. This simple picture anticipates the cancellation found for $C_{3v}$-symmetric trimers in Sec.~\ref{sec:rashba}.

Since the Zeeman coupling in Eq.~\eqref{eq:hamiltonian} acts on the electron spin only, our expansion provides the spin part of $\boldsymbol{\mu}_{\mathrm{tot}}$. The orbital coupling of the field, treated in Sec.~\ref{sec:orbital}, contributes the orbital part, whose leading cubic term is the topological orbital moment $R_{ijk}\chi_{ijk}\hat{\mathbf{n}}_{ijk}$ carried by the ground-state currents of non-coplanar spins even without SOC~\cite{Hoffmann2015,Dias2016,Hanke2016,Dias2017}. It gives rise to an additional energy $-R_{ijk}\chi_{ijk}\,\hat{\mathbf{n}}_{ijk}\cdot\mathbf{B}$, where $\hat{\mathbf{n}}_{ijk}$ is the normal to the plaquette, and is the itinerant counterpart of the orbital-flux mechanism of Mott insulators~\cite{Sen1995,Motrunich2006}. This term is not contained in $\Kv_{ijk}$: without SOC, $\Kv_{ijk}$ vanishes identically, whereas the topological orbital moment does not. Neither of the two terms describes a coupling to an independent orbital degree of freedom. Both are contributions to the same configuration-dependent total moment, obtained by integrating out the electrons for a given spin configuration, and they add, therefore, at linear order in $\mathbf{B}$ without double counting. Double counting would arise only if the same contribution were included twice, for instance by adding $-\mathbf{L}^{\mathrm{TO}}\cdot\mathbf{B}$ to an electronic calculation that already contains the orbital coupling of the field.

Equation~\eqref{eq:hellmann_feynman} clarifies also when the new terms are included in practice. \emph{Ab initio} total-energy calculations performed in an applied Zeeman field contain them automatically, since they rely on the true configuration-dependent moment. In contrast, spin models constructed at zero field and supplemented by the Zeeman energy $-\sum_i\mathbf{M}_i\cdot\mathbf{B}$ of rigid moments, as commonly done, truncate $\boldsymbol{\mu}_{\mathrm{tot}}$ at linear order in the spins and miss these terms.

\section{Orbital coupling of the field and the Mott limit}
\label{sec:orbital}

\subsection{Orbital channel from the ground-state current}

The orbital coupling of the field to the electrons, not included in Eq.~\eqref{eq:hamiltonian}, can be treated within the same framework. To linear order in a uniform field, the Hellmann-Feynman theorem applied to the vector-potential coupling gives $\Omega^{\mathrm{orb}}=-\mathbf{B}\cdot\mathbf{m}^{\mathrm{orb}}_{\mathrm{tot}}(\{\mathbf{S}\})$, where $\mathbf{m}^{\mathrm{orb}}_{\mathrm{tot}}=\frac{1}{2}\int\dd\mathbf{r}\,\mathbf{r}\times\mathbf{j}(\mathbf{r})$ is the orbital moment of the zero-field ground-state current $\mathbf{j}$ in the spin configuration $\{\mathbf{S}\}$. Similarly to the grand potential, the current can be expanded in multiple-scattering processes at the magnetic sites~\cite{Dias2016,FernandesLounis2026}. Since $\mathbf{j}$ is odd under time reversal and the reference system is time-reversal invariant, $\mathbf{j}[\{-\mathbf{S}\}]=-\mathbf{j}[\{\mathbf{S}\}]$ holds exactly, and only terms with an odd number of spins survive, as we verified numerically for the tight-binding triangle of Appendix~\ref{app:microscopic}, with and without SOC. The expansion reads therefore
\begin{equation}
\mathbf{m}^{\mathrm{orb}}_{\mathrm{tot}} = \sum_{k}\mathsf{R}^{(1)}_{k}\,\mathbf{S}_k + \sum_{\langle klm\rangle}\mathbf{R}_{klm}\,\chi_{klm} + \dots ,
\label{eq:morb_expansion}
\end{equation}
where the first term is the conventional orbital moment induced by SOC, while the second one is the chiral orbital moment carried by the ground-state currents of non-coplanar spins even without SOC~\cite{Hoffmann2015,Dias2016,Hanke2016,Dias2017}. The dots denote SOC-induced cubic terms with the tensor structure of Eq.~\eqref{eq:omega_ani}, as well as higher orders. For a planar plaquette without SOC, the currents flow in the plane and $\mathbf{R}_{klm}=R_{klm}\hat{\mathbf{n}}_{klm}$. The orbital channel contributes then
\begin{equation}
\Omega^{\mathrm{orb},\chi}_{ijk} = -\,R_{ijk}\,(\hat{\mathbf{n}}_{ijk}\cdot\mathbf{B})\,\chi_{ijk},
\label{eq:orbital_chi}
\end{equation}
which has the $\cos\theta$ dependence of the orbital-flux term of Sen and Chitra~\cite{Sen1995}, of which it is the itinerant-electron counterpart: the flux threading the plaquette couples to the loop current induced by the non-coplanar spins. Such chirality-induced loop currents are well established in Mott insulators and molecular magnets~\cite{Bulaevskii2008,Trif2008,Kamiya2012} and are at the origin of the topological orbital magnetization of itinerant non-coplanar antiferromagnets such as $\gamma$-FeMn~\cite{Hanke2017}. The total field-linear chirality coupling is thus $(\Kv_{ijk}-R_{ijk}\hat{\mathbf{n}}_{ijk})\cdot\mathbf{B}\,\chi_{ijk}$, where the spin channel requires SOC while the orbital channel does not.

Two remarks are in order. First, in the tight-binding triangle, the chirality-induced loop current, extracted from the $U^{3}$ term and averaged over spin rotations, is finite without SOC, and SOC modifies it only at second order for the $C_{3v}$ geometry. Second, a sizable part of the chiral orbital moment is carried by interatomic currents~\cite{Brinker2020b}. First-principles estimates of $R_{ijk}$ based on moments integrated over atomic cells may therefore underestimate the orbital channel, and the full expression $\frac{1}{2}\int\mathbf{r}\times\mathbf{j}$ should be used instead.

\subsection{Both channels in the Mott limit}

To make contact with the strong-coupling limit, we diagonalized exactly the half-filled Hubbard model on a triangle, with hoppings $-t\,e^{\iu\Phi/3}\sigma_0+\iu\boldsymbol{\nu}_b\cdot\boldsymbol{\sigma}$, an on-site repulsion $U$ and the spin Zeeman coupling $-\mathbf{B}\cdot\boldsymbol{\sigma}$. Projecting the eight lowest states onto the singly occupied subspace provides the effective spin Hamiltonian, from which we extract the coefficient $c_\chi$ of $\mathbf{S}_0\cdot(\mathbf{S}_1\times\mathbf{S}_2)$, with $\mathbf{S}=\boldsymbol{\sigma}/2$. We find the following:

(i) \emph{Orbital channel.} For an orbital flux $\Phi$ without SOC, $|c_\chi|$ converges to the Sen-Chitra coefficient $24t^{3}\sin\Phi/U^{2}$~\cite{Sen1995}, with deviations of 13\%, 3.6\%, 0.9\% and 0.2\% for $U/t=10$, 20, 40 and 80, in line with corrections of order $t^{2}/U^{2}$.

(ii) \emph{Spin channel without SOC.} A spin Zeeman field alone gives $c_\chi=0$ exactly, and so do SOC alone and $B_z$ combined with a $\sigma_z$-conserving SOC. In all these cases, the Zeeman coupling acts on a conserved component of the total spin and commutes with the Hamiltonian, and it cannot generate multi-spin terms. This generalizes the observation of Ref.~\cite{Sen1995}, according to which the Pauli term commutes with the generator of the strong-coupling expansion.

(iii) \emph{Spin channel with SOC.} Rashba-type SOC, $\boldsymbol{\nu}_b=\lambda\,\hat{z}\times\hat{\mathbf{r}}_b$, breaks the conservation of all spin components, and a field-linear chirality coupling emerges. Without field, the SOC enters the chirality coupling of the Hubbard model only in combination with the average spin of the plaquette, i.e., as a four-spin term~\cite{Nikolic2020}. The Zeeman field replaces this additional spin and yields a genuine three-spin coupling. As required by the $C_{3v}$ symmetry, it points along $\hat{z}$ and scales as $\lambda^{2}$: $\partial c_\chi/\partial B_z=-4.8\times10^{-7}$, $-1.9\times10^{-6}$ and $-7.7\times10^{-6}$ for $\lambda/t=0.05$, 0.1 and 0.2 at $U/t=20$. For $t=1$~eV, $\lambda=0.1t$ and a triangle of area 2~\AA$^{2}$, this amounts to about $10^{-10}$~eV per tesla, compared to about $2\times10^{-6}$~eV per tesla for the orbital channel. Here, we used the Zeeman energy $B\approx5.8\times10^{-5}$~eV per tesla and the flux phase $eB_{\mathrm{ext}}A/\hbar\approx3.0\times10^{-5}$ per tesla, where $B_{\mathrm{ext}}$ is the applied field. In the Mott limit with weak SOC, the orbital mechanism dominates thus by orders of magnitude. The spin channel is expected to become competitive in itinerant systems with strong SOC, where the conduction electrons mediating the interactions are strongly spin-orbit coupled. Quantifying this competition requires first-principles calculations of both $\Kv_{ijk}$ and $R_{ijk}$.

(iv) \emph{Additivity of the two channels.} When both couplings of the same field are switched on simultaneously, the chirality coefficient equals the sum of the spin and orbital contributions computed separately, within the numerical accuracy of $10^{-3}$. The two contributions depend on different parameters: the orbital one scales with the flux threading the plaquette and survives without SOC, whereas the spin one scales as $\lambda^{2}$ and vanishes without SOC. The two channels are therefore independent and additive, and their sum involves no double counting.

Recently, local electron spin resonance with a scanning tunneling microscope was proposed to measure the chiral splitting $\delta_\chi=2\mu_\chi B_z$ of spin-$\frac12$ trimers, where $\mu_\chi$ is the chiral orbital moment of the Sen-Chitra term~\cite{FerriCortes2026}. Our results imply that such a splitting contains both channels, $\delta_\chi\propto(\Kv_{ijk}-R_{ijk}\hat{\mathbf{n}}_{ijk})\cdot\mathbf{B}$, with distinct angular dependences. While the orbital contribution vanishes for an in-plane field, the spin-channel contribution of clusters with a symmetry lower than $C_{3v}$ has in-plane components (Sec.~\ref{sec:rashba}). Measuring the chiral splitting as function of the field orientation could therefore disentangle the two mechanisms.

\section{Magnetic atoms in a Rashba electron gas}
\label{sec:rashba}

\subsection{Model and approximations}

To obtain explicit expressions, we consider magnetic atoms embedded in a 1D or 2D electron gas with Rashba SOC~\cite{Rashba1960,Bychkov1984},
\begin{equation}
\mathcal{H}^{0}+\mathcal{H}^{\mathrm{soc}} = -\frac{\hbar^{2}}{2m^{*}}\nabla^{2} + \alpha_R\,(\boldsymbol{\sigma}\times(-\iu\nabla))\cdot\hat{z},
\label{eq:rashba}
\end{equation}
with effective mass $m^{*}$ and Rashba parameter $\alpha_R$. The bands $\epsilon_{\pm}(k)=\hbar^{2}k^{2}/2m^{*}\pm\alpha_R k$ are split by the Rashba wave vector $k_R=m^{*}\alpha_R/\hbar^{2}$. Defining $q=(2m^{*}E/\hbar^{2}+k_R^{2})^{1/2}$, the Green functions read~\cite{Imamura2004,Bouaziz2017}
\begin{align}
\text{1D:}\quad G^{A}_{ij} &= -\iu\frac{m^{*}}{\hbar^{2}}\frac{e^{\iu q r_{ij}}}{q}\cos(k_R r_{ij}), \nonumber\\
\GB_{ij} &= \frac{m^{*}}{\hbar^{2}}\frac{e^{\iu q r_{ij}}}{q}\sin(k_R r_{ij})\,\hat{\boldsymbol{\beta}}_{ij}, \label{eq:G1D}\\
\text{2D:}\quad G^{A}_{ij} &\simeq -\iu\frac{m^{*}}{\hbar^{2}}\frac{e^{\iu(q r_{ij}-\pi/4)}}{\sqrt{2\pi q r_{ij}}}\cos(k_R r_{ij}), \nonumber\\
\GB_{ij} &\simeq \frac{m^{*}}{\hbar^{2}}\frac{e^{\iu(q r_{ij}-\pi/4)}}{\sqrt{2\pi q r_{ij}}}\sin(k_R r_{ij})\,\hat{\boldsymbol{\beta}}_{ij}, \label{eq:G2D}
\end{align}
where $r_{ij}=|\mathbf{r}_{ij}|$, $\mathbf{r}_{ij}=r_{ij}(\cos\beta_{ij},\sin\beta_{ij})$ is the oriented bond vector and $\hat{\boldsymbol{\beta}}_{ij}=(\sin\beta_{ij},-\cos\beta_{ij},0)$ the in-plane unit vector perpendicular to it, with $\hat{\boldsymbol{\beta}}_{ji}=-\hat{\boldsymbol{\beta}}_{ij}$. The 2D expressions are the asymptotic forms, for $qr_{ij}\gg1$ and $k_R\ll q$, of the exact Green functions given in terms of Hankel functions~\cite{Bouaziz2017,Lounis2020}. The magnetic atoms are described by a contact $s$-$d$ interaction $J^{sd}$, and we use the lowest-order (Born, RKKY) approximation $U_i\to J^{sd}$, as in Refs.~\cite{Imamura2004,Lounis2020}. Consistently with Eq.~\eqref{eq:Delta_iso}, we neglect corrections of second order in SOC to the relation $\sum_l G^{A}_{il}G^{A}_{lj}=-\dd G^{A}_{ij}/\dd E$. We denote by $L=r_{ij}+r_{jk}+r_{ki}$ the perimeter and by $P=r_{ij}r_{jk}r_{ki}$ the product of the bond lengths of the trimer, and by $q_F$ the value of $q$ at the Fermi energy.

\subsection{Closed expressions}

The energy integrals are evaluated asymptotically for $q_FL\gg1$ in Appendix~\ref{app:asymptotics}. Since $\dd G_{ij}/\dd q\simeq\iu r_{ij}G_{ij}$ to leading order for both $G^{A}$ and $\GB$, each bond coefficient is proportional to the length of the bond carrying the field vertex. To leading order, we obtain
\begin{equation}
\Delta_{\overline{ij}k}=\frac{r_{ij}}{L}\,\frac{\mathcal{X}_{ijk}}{4},\quad
\Dani^{b'}_{\overline{ij}k} = -\frac{r_{ij}}{L}\,\frac{\mathcal{X}_{ijk}}{4}\,\mathbf{t}_{b'},
\label{eq:rashba_structure}
\end{equation}
with $\mathbf{t}_{b}\equiv\tan(k_R r_{b})\,\hat{\boldsymbol{\beta}}_{b}$,
and where $b'$ labels the bond that carries $\GB$; for $b'=ij$, $\Dani^{b'}_{\overline{ij}k}$ denotes the sum $\Dani^{lj}_{\overline{ij}k}+\Dani^{il}_{\overline{ij}k}$ of Eq.~\eqref{eq:derivative_B}. The total amplitude $\mathcal{X}_{ijk}$ follows from the sum rule, Eq.~\eqref{eq:sum_rule}:
\begin{align}
\mathcal{X}^{\mathrm{1D}}_{ijk} &= -\frac{4}{\pi}\left(\frac{m^{*}J^{sd}}{\hbar^{2}q_F}\right)^{3}\cos(q_FL)\prod_{b}\cos(k_R r_b), \label{eq:X1D}\\
\mathcal{X}^{\mathrm{2D}}_{ijk} &= -\frac{4}{\pi}\left(\frac{m^{*}J^{sd}}{\hbar^{2}}\right)^{3}\frac{\cos(q_FL-\frac{3\pi}{4})}{(2\pi q_F)^{3/2}\sqrt{P}}\prod_{b}\cos(k_R r_b). \label{eq:X2D}
\end{align}
In 1D, Eq.~\eqref{eq:X1D} follows directly from the exact Green functions~\eqref{eq:G1D}, without asymptotic approximation. Combining Eqs.~\eqref{eq:Z_iso}, \eqref{eq:K} and \eqref{eq:rashba_structure} leads to
\begin{align}
\mathcal{Z}_{i(jk)} &= \frac{r_{ij}+r_{ki}-r_{jk}}{L}\,\mathcal{X}_{ijk}, \nonumber\\
\Kv_{ijk} &= \frac{\mathcal{X}_{ijk}}{3}\left[\sum_b\mathbf{t}_b-\frac{4}{L}\sum_b r_b\,\mathbf{t}_b\right],
\label{eq:rashba_results}
\end{align}
where the bonds are oriented along the loop $i\to j\to k\to i$. Equations~\eqref{eq:X1D}--\eqref{eq:rashba_results} constitute the main results for the Rashba model, with the following properties.

(i) \emph{Geometry.} By virtue of the triangle inequality, $\mathcal{Z}_{i(jk)}$ has the sign of $\mathcal{X}_{ijk}$ and vanishes at leading order when $i$ lies between $j$ and $k$ on a line. In a linear trimer, the field couples therefore to the central spin only at next order, with $\mathcal{Z}_{j(ki)}\simeq\frac{2}{\pi}(m^{*}J^{sd}/\hbar^{2})^{3}\prod_b\cos(k_Rr_b)\,\sin(q_FL)/(q_F^{4}r_{ki})$ in 1D (Appendix~\ref{app:asymptotics}).

(ii) \emph{Decay.} Interestingly, the field-induced interactions decay more slowly than their zero-field multi-spin counterparts. The latter scale as $[L\,P^{(d-1)/2}]^{-1}$ in $d$ dimensions~\cite{Lounis2020}, while $\mathcal{X}_{ijk}\propto P^{-(d-1)/2}$. The additional factor $L$ reflects the field-induced change of the Fermi wave vector, $\delta q_F\simeq B/\hbar v_F$, which modifies the phase $q_FL$ of the loop proportionally to its perimeter. In 1D, $\mathcal{X}_{ijk}$ does not decay at all, apart from the Rashba modulation. The linear-in-$B$ description requires consequently $BL/\hbar v_F\ll1$. For the Au(111) surface state, $\hbar v_F\approx4.9$~eV\,\AA{}~\cite{Walls2007,Bouaziz2017}, such that $BL/\hbar v_F\approx0.1$ for clusters of 100~nm at 10~T, and the condition is well satisfied for nanoscale clusters.

(iii) \emph{Chirality coupling.} For $k_R r_b\ll1$, $\mathbf{t}_b\simeq-k_R\,\hat{z}\times\mathbf{r}_b$, and $\sum_b\mathbf{t}_b\simeq0$ since the bond vectors of a closed loop add up to zero. Hence, $\Kv_{ijk}\simeq-(4k_R/3L)\,\mathcal{X}_{ijk}\sum_b r_b^{2}\,\hat{\boldsymbol{\beta}}_b$ is linear in the Rashba coupling and lies in the plane of the electron gas. For Weyl electrons, whose spin is locked to their momentum by an isotropic spin-orbit field, the chirality coupling points instead along the plaquette normal~\cite{Nikolic2021}, which shows that the direction of $\Kv_{ijk}$ is dictated by the form of the SOC. For clusters of low symmetry, the linear in-plane and the second-order out-of-plane contributions coexist and can be of comparable size already for moderate SOC (Appendix~\ref{app:exact}). For an equilateral trimer, the three vectors $\hat{\boldsymbol{\beta}}_b$ are rotated by $120^{\circ}$ with respect to each other, and $\Kv_{ijk}=0$. This is consistent with the symmetry argument of Sec.~\ref{sec:symmetry}, which allows only an out-of-plane $\Kv_{ijk}$ for $C_{3v}$ trimers, a component that cannot be produced by the linear-in-$\alpha_R$ treatment. An explicit microscopic calculation (Appendix~\ref{app:microscopic}) confirms that $\Kv_{ijk}$ of a $C_{3v}$ trimer points along $\hat{z}$ and appears at second order in the Rashba coupling. For a linear trimer with spacing $a$, Eq.~\eqref{eq:rashba_results} gives $\Kv_{ijk}=-\frac{1}{3}\mathcal{X}_{ijk}\tan(2k_Ra)\,\hat{y}$, which points in-plane and perpendicular to the chain, as required by symmetry. The ratio $|\Kv_{ijk}|/|\mathcal{X}_{ijk}|=\frac{1}{3}\tan(2k_Ra)$ reaches about $0.2$ for $a=20$~\AA{} on Au(111).

\begin{figure}[t]
\includegraphics[width=0.95\linewidth]{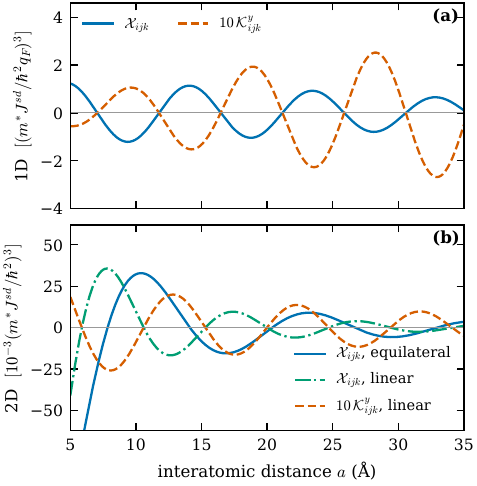}
\caption{Field-induced three-spin coefficients as a function of the interatomic distance $a$, for Rashba parameters of the Au(111) surface state, $q_F=0.167$~\AA$^{-1}$ and $k_R=0.0135$~\AA$^{-1}$~\cite{Walls2007,Bouaziz2017}. (a) 1D linear trimer: total isotropic amplitude $\mathcal{X}_{ijk}$ [Eq.~\eqref{eq:X1D}] and in-plane chirality coupling $\mathcal{K}^{y}_{ijk}$ [Eq.~\eqref{eq:rashba_results}], in units of $(m^{*}J^{sd}/\hbar^{2}q_F)^{3}$. (b) 2D: $\mathcal{X}_{ijk}$ for equilateral and linear trimers [Eq.~\eqref{eq:X2D}], and $\mathcal{K}^{y}_{ijk}$ for the linear trimer, in units of $(m^{*}J^{sd}/\hbar^{2})^{3}$. For the equilateral trimer, $\Kv_{ijk}$ vanishes by symmetry at linear order in the Rashba coupling. The curves are based on the asymptotic expressions, which are accurate for $q_Fa\gtrsim1$.}
\label{fig:rashba}
\end{figure}

Figure~\ref{fig:rashba} shows the resulting coefficients for Au(111) parameters. In 1D, $\mathcal{X}_{ijk}$ oscillates with a period $\pi/2q_F$ as function of the spacing of a linear trimer, its amplitude being modulated only by the Rashba beating $\prod_b\cos(k_Rr_b)$. The chirality coupling $\mathcal{K}^{y}_{ijk}$ oscillates in antiphase with $\mathcal{X}_{ijk}$ for $2k_Ra<\pi/2$, with an amplitude growing as $\tan(2k_Ra)$. In 2D, both coefficients decay as $P^{-1/2}$, and the equilateral and linear geometries differ in phase owing to their different perimeters.

\section{Consequences for magnetic trimers}
\label{sec:trimers}

In the following, we address how the field-induced interactions modify the field response of two model trimers with equal moments $M$, coupled by the bilinear exchange $-J\sum_{\langle nm\rangle}\mathbf{S}_n\cdot\mathbf{S}_m$ with an antiferromagnetic nearest-neighbor coupling $J<0$.

\subsection{Linear Ising trimer: shift of critical fields}

\begin{figure}[t]
\includegraphics[width=0.95\linewidth]{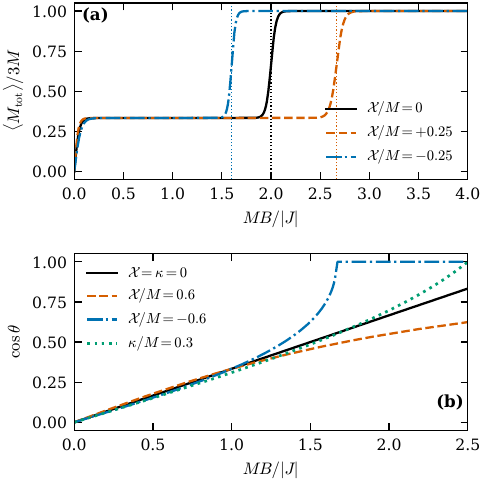}
\caption{Impact of the field-induced three-spin interactions on antiferromagnetic trimers. (a) Magnetization of an Ising linear trimer, Eq.~\eqref{eq:ising_trimer}, at $k_BT=0.05|J|$. Dotted lines mark the critical fields of Eq.~\eqref{eq:Bc}. (b) Canting angle of an equilateral trimer in the $120^{\circ}$ umbrella configuration in a field perpendicular to its plane, obtained by minimizing Eq.~\eqref{eq:umbrella}. The parameters are illustrative and have been chosen large to make the effects visible.}
\label{fig:trimers}
\end{figure}

For an Ising linear trimer (sites $1$-$2$-$3$) in a field along the easy axis, Eq.~\eqref{eq:ising} gives
\begin{equation}
E(\{s\}) = -J\,(s_1s_2+s_2s_3) - MB\sum_n s_n + \mathcal{X}B\,s_1s_2s_3 .
\label{eq:ising_trimer}
\end{equation}
The chirality term plays no role here, since $\chi_{ijk}=0$ for collinear spins. Without exchange, the three-spin term competes with the Zeeman energy on equal footing, both being linear in $B$. The ground state then switches from $(\uparrow\uparrow\uparrow)$ to the threefold degenerate manifold with one reversed spin when $\mathcal{X}>M$, independently of the field strength. For magnetic adatoms, however, $\mathcal{X}$ is of third order in $J^{sd}$, while $M$ amounts to several Bohr magnetons, and this regime is unlikely to be reached. A more relevant effect arises for $J<0$: the low-field ground state $(\uparrow\downarrow\uparrow)$ with total moment $M$ is replaced by the saturated state at the critical field
\begin{equation}
B_c = \frac{2|J|}{M-\mathcal{X}} .
\label{eq:Bc}
\end{equation}
In other words, the three-spin interaction renormalizes the moment that determines the position of the magnetization step. Figure~\ref{fig:trimers}(a) shows the thermal magnetization for several values of $\mathcal{X}/M$. Since critical fields and plateaus in magnetization curves measured by spin-polarized scanning tunneling spectroscopy are routinely used to extract exchange constants of engineered nanostructures~\cite{Meier:2008,Khajetoorians2012}, Eq.~\eqref{eq:Bc} implies that neglecting $\mathcal{X}$ biases such estimates by a relative amount $\mathcal{X}/M$.

\subsection{Equilateral trimer: control of non-coplanarity}

For an equilateral trimer with $J<0$, the zero-field ground state is the coplanar $120^{\circ}$ N\'eel state. A field $\mathbf{B}=B\hat{z}$ perpendicular to the trimer cants the moments into an umbrella configuration characterized by the polar angle $\theta$, the azimuthal angles $0^{\circ}$, $120^{\circ}$ and $240^{\circ}$, and the scalar chirality $\chi_{ijk}=\frac{3\sqrt{3}}{2}\sin^{2}\theta\cos\theta$. With $c\equiv\cos\theta$ and $\mathcal{Z}_{i(jk)}=\mathcal{X}/3$ for all sites, one finds $\mathbf{S}_i\cdot\mathbf{S}_j=(3c^{2}-1)/2$, and the energy reads
\begin{align}
E(c) &= -\frac{3J}{2}\,(3c^{2}-1) - 3MB\,c \nonumber\\
&\quad + \frac{\mathcal{X}}{2}\,B\,(3c^{2}-1)\,c + \kappa\,B\,c\,(1-c^{2}).
\label{eq:umbrella}
\end{align}
The last term collects the relativistic contributions. If all vectors $\Dani$ are parallel to $\hat{z}$, as allowed by the $C_{3v}$ symmetry, every bracket of Eq.~\eqref{eq:omega_ani} is proportional to $\sin^{2}\theta\cos\theta$, and $\kappa$ is a linear combination of their $z$ components. For weak fields, minimizing the energy gives
\begin{equation}
\cos\theta \simeq \frac{MB}{3|J|}\left(1+\frac{\mathcal{X}}{6M}-\frac{\kappa}{3M}\right).
\label{eq:canting}
\end{equation}
Depending on their signs, the isotropic and relativistic three-spin terms enhance or reduce the field-induced non-coplanarity, and therefore the chirality $\chi_{ijk}$, which is the relevant quantity for the topological Hall response. On their own, the two terms favor $\cos\theta=\pm1/3$ and $\cos^{2}\theta=1/3$, respectively. Interestingly, Fig.~\ref{fig:trimers}(b) shows that a sufficiently negative $\mathcal{X}$ can drive a discontinuous transition to the saturated state. As discussed in Sec.~\ref{sec:rashba}, $\kappa$ vanishes for this geometry in the linear-in-$\alpha_R$ Rashba model, and a first-principles evaluation for real surfaces is needed to quantify it.

\subsection{Implications for skyrmions}
\label{sec:skyrmions}

In extended two-dimensional magnets, the field-induced interactions take simple continuum forms. For a slowly varying magnetization of fixed modulus and triangles ordered counterclockwise, the exact identity $\chi_{ijk}=\mathbf{S}_i\cdot[(\mathbf{S}_j-\mathbf{S}_i)\times(\mathbf{S}_k-\mathbf{S}_i)]$ gives $\chi_{\triangle}\simeq2A_{\triangle}\,\mathbf{m}\cdot(\partial_x\mathbf{m}\times\partial_y\mathbf{m})$. On a triangular lattice, each sublattice of triangles has the density $1/2A_{\triangle}$, such that the lattice prefactors cancel and
\begin{equation}
E_{\mathcal{K}} \simeq \int\!\dd^{2}r\;\kappa(\mathbf{r})\,\mathbf{m}\cdot(\partial_x\mathbf{m}\times\partial_y\mathbf{m}) = 4\pi\kappa\,Q ,
\label{eq:EK_Q}
\end{equation}
where $\kappa=(\Kv_{\blacktriangle}+\Kv_{\blacktriangledown})\cdot\mathbf{B}$ collects the two inequivalent triangles of a (111) surface and $Q$ is the topological charge. For uniform $\kappa$, this term leaves the skyrmion profile unchanged, splits skyrmions from antiskyrmions by $8\pi|\kappa|$, and exerts a force $-4\pi Q\,\nabla\kappa$ when $\kappa$ varies in space. The relativistic terms of Eq.~\eqref{eq:omega_ani} that are not proportional to $\chi_{ijk}$ have a different continuum limit: for a perpendicular field, they reduce to a field-dependent interfacial DMI, $D\to D+\delta D\,m_z$. In the Rashba model, lattice sums for Néel skyrmions give $\delta\mathbf{D}_{ij}=\frac{4}{3}\tan(k_Ra)\,\mathcal{X}_\triangle B_z\,\hat{\boldsymbol{\beta}}_{ij}$ for nearest-neighbor bonds. In contrast to $\mathcal{K}_z$, this term is linear in SOC and survives for $C_{3v}$ interfaces. Compared to the zero-field DMI of the same model, $\delta D/D\sim(m^{*}J^{sd}/\hbar^{2})\,B/E_F$, which is of the order of $10^{-3}$ at 10~T even for strong coupling and becomes appreciable only close to a sign change of $D$. Finally, these statements assume a rigid modulus. For the soft moduli of itinerant magnets~\cite{Rossler2006}, $\kappa$ depends on the local amplitude of the magnetization, and $E_{\mathcal{K}}$ is no longer purely topological. A quantitative assessment of these effects requires first-principles values of $\Kv_{ijk}$ and $\mathcal{X}_{ijk}$, which is left for future work.

\section{Conclusion}
\label{sec:conclusion}

In this work, we derived analytically how the Zeeman coupling of conduction electrons generates three-spin interactions linear in the magnetic field. These are the lowest-order corrections to the Zeeman energy allowed by time-reversal symmetry, and they are absent from the spin models commonly used to describe magnets in a field. Utilizing a Green-function expansion of the grand potential, we identified two classes of interactions. The isotropic ones, $\mathcal{Z}_{i(jk)}(\mathbf{S}_j\cdot\mathbf{S}_k)(\mathbf{S}_i\cdot\mathbf{B})$, do not require SOC. They stem from the field-induced polarization of the conduction electrons, have no counterpart in SU(2)-symmetric Mott insulators, where the Zeeman coupling acts on the conserved total spin, and their total strength $\mathcal{X}_{ijk}$ is fixed by a Fermi-surface sum rule. The relativistic interactions include the coupling $\Kv_{ijk}\cdot\mathbf{B}\,\chi_{ijk}$ of the scalar spin chirality to the field. We demonstrated that this coupling emerges from processes in which one electron propagator carries the SOC-induced spin dependence, such that $\Kv_{ijk}$ vanishes without SOC, and we provided an explicit expression for $\Kv_{ijk}$. This mechanism is complementary to the orbital-flux mechanism known from Mott insulators~\cite{Sen1995,Motrunich2006}, which we recovered within the same framework as the Zeeman energy of the chirality-induced ground-state current. Exact diagonalization of the Hubbard trimer reproduces the Sen-Chitra coefficient and shows that, in the Mott limit, the spin channel requires a SOC breaking the conservation of the spin component along the field. The mechanism differs from the field-induced three-spin term of the Kitaev model, which is cubic in the field~\cite{Kitaev2006}, and represents a three-spin, field-enabled counterpart of the zero-field four-spin chiral interactions~\cite{Grytsiuk2020,Brinker2020}.

For magnetic atoms embedded in Rashba electron gases, we obtained closed expressions for all coefficients to leading order in SOC. The field-induced interactions decay more slowly than the zero-field multi-spin interactions, since the field acts on the phase accumulated along the whole perimeter of the cluster. The chirality coupling is linear in the Rashba parameter and oriented in-plane. It cancels for $C_{3v}$-symmetric trimers, while it survives for clusters of lower symmetry such as linear trimers. In $C_{3v}$ trimers, an out-of-plane coupling appears only at second order in the Rashba coupling. For antiferromagnetic trimers, the isotropic term renormalizes the critical fields of magnetization steps, and both classes of interactions modify the field-induced canting and thus the scalar chirality. In skyrmion hosts, the chirality coupling reduces in the continuum limit to an energy $4\pi\kappa Q$, proportional to the topological charge, which splits skyrmions from antiskyrmions, while the remaining relativistic terms produce a field-dependent DMI, $D\to D+\delta D\,m_z$, of relative magnitude $\sim(m^{*}J^{sd}/\hbar^{2})B/E_F$, linear in SOC and enhanced close to a sign change of $D$.

Our findings open several perspectives. First-principles evaluations of Eqs.~\eqref{eq:Z_iso} and \eqref{eq:K}, for instance with the Korringa-Kohn-Rostoker Green-function method, are needed to quantify these interactions in real materials. Since the interactions stem from the field-induced polarization of the host, whose strength is controlled by the ratio of the pair exchange to the Fermi energy, they are expected to be small for weakly polarizable hosts such as Cu(111). Promising candidates are strongly polarizable hosts close to the Stoner instability, such as Pt, Pd and Rh, as well as itinerant magnets with non-coplanar order, where Hall effects driven by the scalar chirality have been observed~\cite{Machida2010,Takagi2023}, and kagome magnets deposited on heavy-metal substrates~\cite{Fujiwara2021}. With such input, atomistic spin simulations could quantify the impact on the stability and dynamics of skyrmions. First-principles calculations of both the spin-channel coupling $\Kv_{ijk}$ and the orbital-channel coefficient $R_{ijk}$, including interatomic currents, would establish which mechanism dominates in a given material. Finally, since $\Kv_{ijk}\cdot\mathbf{B}\,\chi_{ijk}$ depends on the direction of the field with respect to the crystal, we anticipate that the orientation of an applied field could be utilized to control non-coplanar spin-textures and the associated topological transport.

\begin{acknowledgments}
This work was supported by the Deutsche Forschungsgemeinschaft (DFG) through Project No.~328545488 -- CRC/TRR 227, Project No.~B12, and by the German Excellence Strategy -- EXC 3112/1 -- 533767171 (Center for Chiral Electronics). The author discloses the use of AI tools to assist with drafting and refining portions of the manuscript based on author-provided scientific ideas and context that were critically reviewed, verified, and revised.
\end{acknowledgments}

\section*{Data availability}
All data shown in the figures can be reproduced from the analytical expressions given in the text.

\appendix

\section{Pauli-trace identities}
\label{app:traces}

With $\sigma_{\mathbf{a}}\equiv\mathbf{a}\cdot\boldsymbol{\sigma}$, the isotropic terms follow from
\begin{align}
&\tfrac{1}{2}\Tr_s\left[\sigma_{\mathbf{S}_j}\sigma_{\mathbf{S}_k}\sigma_{\mathbf{S}_i}\sigma_{\mathbf{B}}\right] \nonumber\\
&= (\mathbf{S}_j\cdot\mathbf{S}_k)(\mathbf{S}_i\cdot\mathbf{B}) - (\mathbf{S}_j\cdot\mathbf{S}_i)(\mathbf{S}_k\cdot\mathbf{B}) \nonumber\\
&\quad + (\mathbf{S}_j\cdot\mathbf{B})(\mathbf{S}_k\cdot\mathbf{S}_i).
\end{align}
The relativistic terms follow from traces of five Pauli matrices, which are purely imaginary for real vectors. For an arbitrary vector $\mathbf{D}$,
\begin{widetext}
\begin{align}
\tfrac{1}{2\iu}\Tr_s\left[\sigma_{\mathbf{S}_j}\sigma_{\mathbf{S}_k}\sigma_{\mathbf{D}}\sigma_{\mathbf{S}_i}\sigma_{\mathbf{B}}\right] &= \mathbf{D}\cdot\left[-\chi_{ijk}\mathbf{B} + (\mathbf{S}_i\times\mathbf{B})(\mathbf{S}_j\cdot\mathbf{S}_k) + (\mathbf{S}_j\times\mathbf{S}_k)(\mathbf{S}_i\cdot\mathbf{B}) + \mathbf{S}_i\,(\mathbf{S}_j\times\mathbf{S}_k)\cdot\mathbf{B}\right],\\
\tfrac{1}{2\iu}\Tr_s\left[\sigma_{\mathbf{S}_j}\sigma_{\mathbf{D}}\sigma_{\mathbf{S}_k}\sigma_{\mathbf{S}_i}\sigma_{\mathbf{B}}\right] &= \mathbf{D}\cdot\left[-\chi_{ijk}\mathbf{B} - (\mathbf{S}_j\times\mathbf{B})(\mathbf{S}_k\cdot\mathbf{S}_i) + (\mathbf{S}_k\times\mathbf{S}_i)(\mathbf{S}_j\cdot\mathbf{B}) + \mathbf{S}_j\,(\mathbf{S}_k\times\mathbf{S}_i)\cdot\mathbf{B}\right],\\
\tfrac{1}{2\iu}\Tr_s\left[\sigma_{\mathbf{D}}\sigma_{\mathbf{S}_j}\sigma_{\mathbf{S}_k}\sigma_{\mathbf{S}_i}\sigma_{\mathbf{B}}\right] &= \mathbf{D}\cdot\left[+\chi_{ijk}\mathbf{B} + (\mathbf{S}_j\times\mathbf{B})(\mathbf{S}_k\cdot\mathbf{S}_i) - (\mathbf{S}_k\times\mathbf{S}_i)(\mathbf{S}_j\cdot\mathbf{B}) + \mathbf{S}_j\,(\mathbf{S}_k\times\mathbf{S}_i)\cdot\mathbf{B}\right],\\
\tfrac{1}{2\iu}\Tr_s\left[\sigma_{\mathbf{S}_j}\sigma_{\mathbf{S}_k}\sigma_{\mathbf{S}_i}\sigma_{\mathbf{D}}\sigma_{\mathbf{B}}\right] &= \mathbf{D}\cdot\left[+\chi_{ijk}\mathbf{B} - (\mathbf{S}_i\times\mathbf{B})(\mathbf{S}_j\cdot\mathbf{S}_k) - (\mathbf{S}_j\times\mathbf{S}_k)(\mathbf{S}_i\cdot\mathbf{B}) + \mathbf{S}_i\,(\mathbf{S}_j\times\mathbf{S}_k)\cdot\mathbf{B}\right].
\end{align}
\end{widetext}
These identities correspond to $\mathbf{G}^{B}$ placed on the segments $ki$, $jk$, $lj$ and $il$, respectively. Averaging the four brackets over global rotations $\mathbf{S}_n\to R\mathbf{S}_n$, $R\in\mathrm{SO}(3)$, at fixed $\mathbf{B}$ projects them onto their rotation-invariant parts, $-\frac{1}{3}\chi_{ijk}\mathbf{B}$, $-\frac{1}{3}\chi_{ijk}\mathbf{B}$, $\chi_{ijk}\mathbf{B}$ and $\chi_{ijk}\mathbf{B}$, respectively, using $\langle R\mathbf{a}\,(R\mathbf{b})^{\mathsf{T}}\rangle_R=\frac{1}{3}(\mathbf{a}\cdot\mathbf{b})\,\mathbb{1}$. This yields Eq.~\eqref{eq:K}. The averages have been verified numerically using the icosahedral group, which averages polynomials of degree three in $R$ exactly. For the loop of opposite orientation, reversing the order of the matrices complex-conjugates the trace, while $\mathbf{G}^{B}_{ij}=-\mathbf{G}^{B}_{ji}$ contributes an additional sign. The two orientations therefore contribute equally, which gives the prefactor $4$ in Eqs.~\eqref{eq:Z_iso} and \eqref{eq:omega_ani}, together with the factor $2$ from the spin trace.

\section{Asymptotic evaluation}
\label{app:asymptotics}

At zero temperature, the energy integrals are converted into integrals over $q$ with upper limit $q_F$. They are evaluated by repeated integration by parts,
\begin{equation}
\int^{q_F}\!\dd q\,e^{\iu qL}g(q) = e^{\iu q_FL}\left[\frac{g(q_F)}{\iu L}+\frac{g'(q_F)}{L^{2}}+\mathcal{O}(L^{-3})\right],
\end{equation}
which is equivalent to using the asymptotic expansions of the sine, cosine and Fresnel integrals. In 1D, Eq.~\eqref{eq:G1D} gives $\dd G^{A}_{ij}/\dd q=(m^{*}/\hbar^{2})\cos(k_Rr_{ij})\,e^{\iu qr_{ij}}(r_{ij}+\iu/q)/q$, and Eq.~\eqref{eq:Delta_iso} becomes
\begin{align}
\Delta_{\overline{ij}k} &= \frac{1}{\pi}\left(\frac{m^{*}J^{sd}}{\hbar^{2}}\right)^{3}\prod_b\cos(k_Rr_b) \nonumber\\
&\quad\times\mathrm{Im}\!\int^{q_F}\!\dd q\,e^{\iu qL}\left(\frac{r_{ij}}{q^{3}}+\frac{\iu}{q^{4}}\right).
\end{align}
To leading order this yields Eq.~\eqref{eq:rashba_structure}. At the next order,
\begin{align}
\mathcal{Z}_{i(jk)} &\simeq -\frac{4}{\pi}\left(\frac{m^{*}J^{sd}}{\hbar^{2}}\right)^{3}\prod_b\cos(k_Rr_b) \nonumber\\
&\quad\times\left[\rho\,\frac{\cos(q_FL)}{q_F^{3}}+(3\rho-1)\frac{\sin(q_FL)}{q_F^{4}L}\right],
\end{align}
with $\rho=(r_{ij}+r_{ki}-r_{jk})/L$.
Summing over the three sites, $\sum\rho=1$ and $\sum(3\rho-1)=0$, which reproduces the exact result~\eqref{eq:X1D} and provides a consistency check of the expansion. For a linear trimer with central site $j$, $\rho=0$ and $L=2r_{ki}$, which gives the expression quoted in Sec.~\ref{sec:rashba}. The relativistic coefficients follow in the same way, since $\dd\GB_{ij}/\dd q\simeq\iu r_{ij}\GB_{ij}$. Replacing $G^{A}$ on bond $b'$ by $\GB$ amounts to $\cos(k_Rr_{b'})\to\sin(k_Rr_{b'})\hat{\boldsymbol{\beta}}_{b'}$ together with a factor $\iu$, and the real part in Eq.~\eqref{eq:Delta_ani} converts this into the relation~\eqref{eq:rashba_structure}. In 2D, the phase factor $e^{-3\iu\pi/4}$ of three propagators and the factor $(2\pi q)^{-3/2}P^{-1/2}$ lead to Eq.~\eqref{eq:X2D}.

\section{Microscopic check on a tight-binding trimer}
\label{app:microscopic}

To test the symmetry statements independently of the multiple-scattering expansion, we consider electrons hopping on a triangle with hoppings $-t_b\sigma_0+\iu\boldsymbol{\nu}_b\cdot\boldsymbol{\sigma}$, local couplings $U\,\mathbf{S}_n\cdot\boldsymbol{\sigma}$ to three classical moments, and two electrons in the lowest, nondegenerate shell. The Rashba-type SOC is modelled by $\boldsymbol{\nu}_b=\lambda\,t_b r_b\,\hat{z}\times\hat{\mathbf{r}}_b$. By Eq.~\eqref{eq:hellmann_feynman}, the chirality coupling equals minus the rotation-invariant part of the zero-field spin moment of the electrons per unit chirality. We extract it from the $U^{3}$ term of the spin moment, averaged exactly over global spin rotations. The results are:
(i) Without SOC, $\Kv$ vanishes to machine precision, for any geometry.
(ii) For an equilateral triangle with equal hoppings, $\Kv\parallel\hat{z}$ and $\mathcal{K}_z\propto\lambda^{2}$. The in-plane components vanish, in agreement with the $C_{3v}$ cancellation at linear order.
(iii) For an isosceles triangle, the $C_s$ symmetry allows $\mathcal{K}_y$ and $\mathcal{K}_z$. A component $\mathcal{K}_y\propto\lambda$ appears when the three bonds are inequivalent, which we model by $t_b\propto e^{-1.5(r_b-a)/a}$, in addition to $\mathcal{K}_z\propto\lambda^{2}$. In the continuum model, the bonds are distinguished by the field-induced phase $\propto r_b$, which yields the linear term of Eq.~\eqref{eq:rashba_results}.

\section{Exact test on a tight-binding metal}
\label{app:exact}

We tested Eqs.~\eqref{eq:Z_iso}, \eqref{eq:sum_rule} and \eqref{eq:K} against exact calculations for a metallic host. The host is a square-lattice tight-binding model with nearest-neighbor hopping $t$, chemical potential $\mu=-t$ and temperature $T=0.02t$--$0.04t$. Where present, it includes a Rashba term $\lambda(\sigma_x\sin k_y-\sigma_y\sin k_x)$. Three magnetic sites form a scalene triangle at $(0,0)$, $(2a,0)$ and $(0,3a)$. The grand potential is obtained from Lloyd's formula, $\Omega=\frac{1}{\pi}\mathrm{Im}\int\dd E\,f\,\ln\det(1-G_{\mathbf{B}}V)$. Here, the Zeeman field is included exactly in the host Green function $G_{\mathbf{B}}$, and $V$ is the local exchange on the three sites. The part trilinear in $U_i,U_j,U_k$ and linear in $\mathbf{B}$ is extracted by finite differences. Energy integrals are evaluated as Matsubara sums. Table~\ref{tab:exact} summarizes the results.
(i) The three isotropic coefficients, which are all different for this geometry, agree with Eq.~\eqref{eq:Z_iso} to better than $3\times10^{-4}$, the accuracy of the finite differences. Fitting the exact energies of eight random configurations requires no invariants beyond those of Eq.~\eqref{eq:iso_invariants}; the residual is $4\times10^{-9}$.
(ii) The sum rule, Eq.~\eqref{eq:sum_rule}, is satisfied to the same accuracy. A direct evaluation of the Fermi-surface loop on the real axis, with a broadening of $0.01t$, agrees within 3\%.
(iii) Beyond weak coupling, the effective coefficient $\mathcal{X}(U)/U^{3}$ of collinear configurations deviates from the leading-order value by 0.1\%, 0.7\%, 2.5\%, 8\% and 75\% for $U/t=0.05$, 0.2, 0.5, 1 and 2, respectively. The leading-order expressions are therefore quantitative for $U\lesssim t$, i.e., up to about one eighth of the bandwidth. Beyond that, $U_i$ should be replaced by the scattering matrices $t_i$.
(iv) With Rashba SOC, the exact chirality coupling, obtained by averaging over global spin rotations, agrees with Eq.~\eqref{eq:K} to 2.6\% for $\lambda=0.1t$ and to 9.5\% for $\lambda=0.2t$. The deviation scales as $\lambda^{2}$. The projection weights matter: replacing the weights $-\frac{1}{3}$ by $-1$, i.e., keeping only the terms explicitly proportional to $\chi_{ijk}\mathbf{B}$, overestimates $\Kv_{ijk}$ by about a factor of two. The exact calculation also yields an out-of-plane component $\mathcal{K}_z\propto\lambda^{2}$, which is beyond the linear order of Eq.~\eqref{eq:K}. For this low-symmetry triangle, it is already comparable to the in-plane components at $\lambda=0.1t$.

\begin{table}[t]
\caption{Exact (Lloyd formula) versus analytical coefficients for the scalene triangle of Appendix~\ref{app:exact}, in units of $t^{-2}$ per $U_iU_jU_k$. $\Kv$ is given for $\lambda=0.1t$.}
\label{tab:exact}
\small
\begin{ruledtabular}
\begin{tabular}{lcc}
Quantity & Analytical & Exact \\
\hline
$\mathcal{Z}_{i(jk)}$ & $9.928\times10^{-4}$ & $9.929\times10^{-4}$ \\
$\mathcal{Z}_{j(ki)}$ & $5.879\times10^{-4}$ & $5.879\times10^{-4}$ \\
$\mathcal{Z}_{k(ij)}$ & $8.364\times10^{-4}$ & $8.366\times10^{-4}$ \\
$\mathcal{X}=\sum\mathcal{Z}$ & $2.4171\times10^{-3}$ & $2.4174\times10^{-3}$ \\
$(\mathcal{K}_x,\mathcal{K}_y)$, Eq.~\eqref{eq:K} & $(2.21,1.28)\times10^{-5}$ & $(2.15,1.24)\times10^{-5}$ \\
$(\mathcal{K}_x,\mathcal{K}_y)$, weights $-1$ & $(4.36,1.89)\times10^{-5}$ & -- \\
$\mathcal{K}_z$ & $0$ (linear order) & $1.17\times10^{-5}$ \\
\end{tabular}
\end{ruledtabular}
\end{table}

\section{Rashba bands in a Zeeman field}
\label{app:bands}

We treat the field to linear order. This is justified by the exact dispersion of the 2D Rashba gas in a perpendicular Zeeman field,
\begin{equation}
E_{\pm}(k) = \frac{\hbar^{2}k^{2}}{2m^{*}} \pm \sqrt{\alpha_R^{2}k^{2}+B^{2}} .
\end{equation}
The field opens a gap $2B$ at $k=0$. For $B<2E_{\mathrm{so}}$, with $E_{\mathrm{so}}=\hbar^{2}k_R^{2}/2m^{*}$, the lower band has a local maximum at $k=0$ and a ring of minima at $k_{\mathrm{so}B}=k_R\sqrt{1-(B/2E_{\mathrm{so}})^{2}}$. Near the Fermi energy, $\alpha_Rk_F\gg B$ and $E_{\pm}\simeq\hbar^{2}k^{2}/2m^{*}\pm\alpha_Rk\,[1+B^{2}/(2\alpha_R^{2}k^{2})]$, so that the field changes the Rashba bands only weakly, and an expansion in $B/E_F$ and $B/\alpha_Rk_F$ is well controlled. For Au(111), $E_{\mathrm{so}}\approx2.7$~meV, and the band topology is preserved up to fields of the order of $90$~T. Together with the condition $BL/\hbar v_F\ll1$ discussed in Sec.~\ref{sec:rashba}, this justifies the linear-response treatment for all realistic laboratory fields and cluster sizes.

\bibliography{references}

\end{document}